\documentstyle[aas2pp4]{article}

\lefthead{Shirey, Bradt, \& Levine}
\righthead{The Complete ``Z'' Track of Cir X-1}

\newcommand{\RXTE}{{\it RXTE\,}}
\newcommand{\EXOSAT}{{\it EXOSAT\,}}
\newcommand{\ASCA}{{\it ASCA\,}}
\newcommand{\ROSAT}{{\it ROSAT\,}}
\newcommand{\Ginga}{{\it Ginga\,}}

\newcommand{\simlt}{{\stackrel{<}{_\sim}}}
\newcommand{\simgt}{{\stackrel{>}{_\sim}}}
\newcommand{\msun}{${\rm M}_\odot$}

\begin{document}

\title{The Complete ``Z'' Track of Circinus X-1} 

\author{
Robert E. Shirey\altaffilmark{1,2}, 
Hale V. Bradt\altaffilmark{1}, and
Alan M. Levine}

\affil{
Center for Space Research, Massachusetts Institute of Technology, Cambridge, MA 02139 
\\
{\bf To be published in the May 20, 1999 issue of The Astrophysical Journal (Vol. 517)}
}

\altaffiltext{1}{Department of Physics, Massachusetts Institute of
Technology, Cambridge, MA 02139}

\altaffiltext{2}{
Present address: Department of Physics, University of California, Santa Barbara,
Santa Barbara, CA 93106; shirey@orion.physics.ucsb.edu}

\begin{abstract}

We carried out an extensive {\em Rossi X-ray Timing Explorer}
campaign, in 1997 June, to study Circinus~X-1 during the active
portion of its 16.55-d intensity cycle.  The observations spanned 10
days, including 56\% coverage for 7~d, and allowed us to find time
segments which clearly demonstrate continuous evolution along the
horizontal, normal, and flaring branches (HB/NB/FB) of a Z-source
low-mass X-ray binary. These results confirm and extend the behavior
we inferred from earlier observations. Here we study the continuous
evolution of the Fourier power spectra and the energy spectra around
the complete hardness-intensity track.

A narrow quasi-periodic oscillation (QPO) peak, previously observed in
the power spectra at 1.3--32~Hz, increases in frequency from 12~Hz to
25~Hz moving down a vertical extension of the horizontal branch in the
hardness-intensity diagram. These horizontal branch QPOs (HBOs) occur
near 30~Hz and fade in strength on the horizontal portion of the HB,
while a broad peak in the power spectrum arises near 4~Hz. This peak
becomes much more prominent along the normal branch and remains near
4~Hz (the normal branch QPOs, or NBOs). On the flaring branch, neither
QPO is present and the power spectrum is dominated by very low
frequency noise.
We also found that each branch of the spectral track is associated
with a specific type of evolution of the energy spectrum. We explored
various models for the energy spectrum and parameterized the evolution
of the spectrum in terms of a two-component model consisting of a
multi-temperature ``disk blackbody'' and a higher-temperature
($\sim$2~keV) blackbody.  We also show that an unusual line-
or edge-like feature occurs at about 10~keV in energy spectra from the
flaring branch and lower normal branch. This unusual feature is very
similar to one seen on the FB and lower NB of the Z~source GX~5$-$1.

\end{abstract}


\section{Introduction}
\label{sec:intro}
Circinus~X-1 exhibits a dramatic 16.55-d cycle of X-ray flaring which
is believed to be the result of enhanced mass transfer occurring near
periastron of a highly eccentric binary orbit (\cite{kaluzienski76};
\cite{murdin80}).  Although the identification of the orbital period
seems to be secure, there is very little direct evidence concerning
the masses of the two components of the binary and the other orbital
parameters. Nonetheless, the compact component is thought to be a
weakly magnetized neutron star on the basis of three type~I X-ray
bursts seen with \EXOSAT\ (\cite{tennant86}).  Further type~I bursts
have not been observed from Cir X-1 since the \EXOSAT\ discovery,
possibly because the source intensity has been higher during
subsequent observations.  No coherent pulsations, which would be
expected to be present if the compact star is a strongly magnetized
neutron star, have been detected (\cite{dower82}; \cite{vaughan94}).

Quasi-periodic oscillations (QPOs) were seen at 1.4~Hz, 5--20~Hz,
and 100--200~Hz in \EXOSAT\ observations of Cir~X-1 in a bright state
(\cite{tennant87}; \cite{tennant88}), but other observations at lower
intensity showed no such QPOs (\cite{oosterbroek95}). Based on these
data, it was suggested that Cir~X-1 is an atoll-source low-mass X-ray
binary (LMXB) that can uniquely reach the Eddington accretion rate and
exhibit normal/flaring branch QPOs (NBOs/FBOs) at 5--20~Hz
(\cite{oosterbroek95}; \cite{klis94}). Similar QPOs were observed in
\Ginga\ observations of Cir~X-1 (\cite{makino93}).

Observations with the All-Sky Monitor (ASM) and Proportional Counter
Array (PCA) on the {\em Rossi X-ray Timing Explorer} (\RXTE) have
shown that since early 1996 (from the beginning of \RXTE\ monitoring)
the baseline intensity level of Cir~X-1 has remained bright
($\sim$1~Crab, 2--12~keV; $\sim$1060~$\mu$Jy at 5.2~keV) at all phases
of the 16.55-d cycle, with both dips and flares associated with phase
zero of the cycle (Shirey et~al.\ 1996\nocite{shirey96},
1998\nocite{shirey98:feb97}, hereafter Papers I \& II;
\cite{shirey98:phd}).  During non-flaring phases (intensity $\approx$
1~Crab) of a cycle in 1996 March, the centroid frequency of a narrow
QPO was observed to evolve between 1.3~Hz and 12~Hz and was strongly
correlated with the cutoff frequency of low-frequency noise and with
the centroid frequency of a broad peak ranging from 20--100~Hz, at
$\sim$13 times the frequency of the lower frequency QPO
(\cite{shirey96})\@. During portions of a more active cycle in 1997
February--March, a similar narrow QPO evolved between 6.8--32~Hz,
while during other portions of the cycle, a broad QPO was observed at
4~Hz (\cite{shirey98:feb97})\@.  By correlating the timing properties
with fragmented branches in hardness-intensity diagrams, we identified
horizontal, normal, and flaring branches. Thus, we interpreted the
narrow 1.3--32~Hz QPOs in Cir~X-1, including the 5--20~Hz QPOs
observed with \EXOSAT, as horizontal-branch oscillations (HBOs), and
the broad 4~Hz QPOs as NBOs, thus calling into question the earlier
interpretation of the QPOs in Cir~X-1.

In this paper we present results from a 10-day set of high-efficiency
\RXTE\ observations, including 56\% coverage for 7~d starting 1.5~d
before phase zero. Portions of these data show a complete Z-source
track for Cir~X-1. These results confirm our previous interpretation
based on incomplete and shifted spectral tracks constructed from
shorter and more widely spaced observations.  We show how both the
Fourier power spectra and the energy spectra evolve along the various
spectral tracks.  We explore possible models for the energy spectrum
and parameterize the evolution by fitting the data at various points
along the hardness-intensity track to a model consisting of a
multi-temperature ``disk blackbody'' and a higher-temperature
isothermal blackbody.  

\section{Observations} 

The PCA light curves and a hardness ratio (``broad color'') for the
1997 June observations are shown in Figure~\ref{fig:june97_10d}. These
data show only moderate variability for the first day of
observations. The source entered a phase of significant dipping during
the half day before phase zero (day 611.5), based on the radio
ephemeris (\cite{stewart91}). The hardness ratio shows that dramatic
spectral evolution, both hardening and softening, occurs during these
dips.

By phase zero the main dipping episode ends, and the transition to the
flaring state begins. While the intensity increases by more than a
factor of three in the lowest energy band (2--6.3~keV), the intensity
between 6.3~keV and 13~keV does not climb at all, and above 13~keV it
decreases by a factor of about 10 over the first 1.5~days following
phase zero. This anti-correlation of the low and high-energy intensity
during the transition results in a decreased hardness ratio after
phase zero, as is observed in \RXTE\ ASM data (Papers
I\nocite{shirey96} \& II\nocite{shirey98:feb97};
\cite{shirey98:phd}). 

After a relatively smooth transition toward high total intensity
(2--18~keV) during the first day following phase zero, the intensity
becomes highly variable (i.e., the ``active'' or ``flaring'' state)
for the remaining 7 days of the observation. The intensity in the hard
band remains much lower than before phase zero but shows strong
variability with a peak value that gradually increases from about 24 to
36 counts~s$^{-1}$~PCU$^{-1}$.  Based on the results in
\cite{shirey98:feb97} and those discussed below, we can identify
short-term variability within restricted limits as motion along
branches in hardness-intensity diagrams and the gradual evolution of
the limits as shifting of the branches.

\section{Complete Track in Hardness-intensity and Color-color Diagrams}

\begin{figure}
\epsscale{1}
\plotone{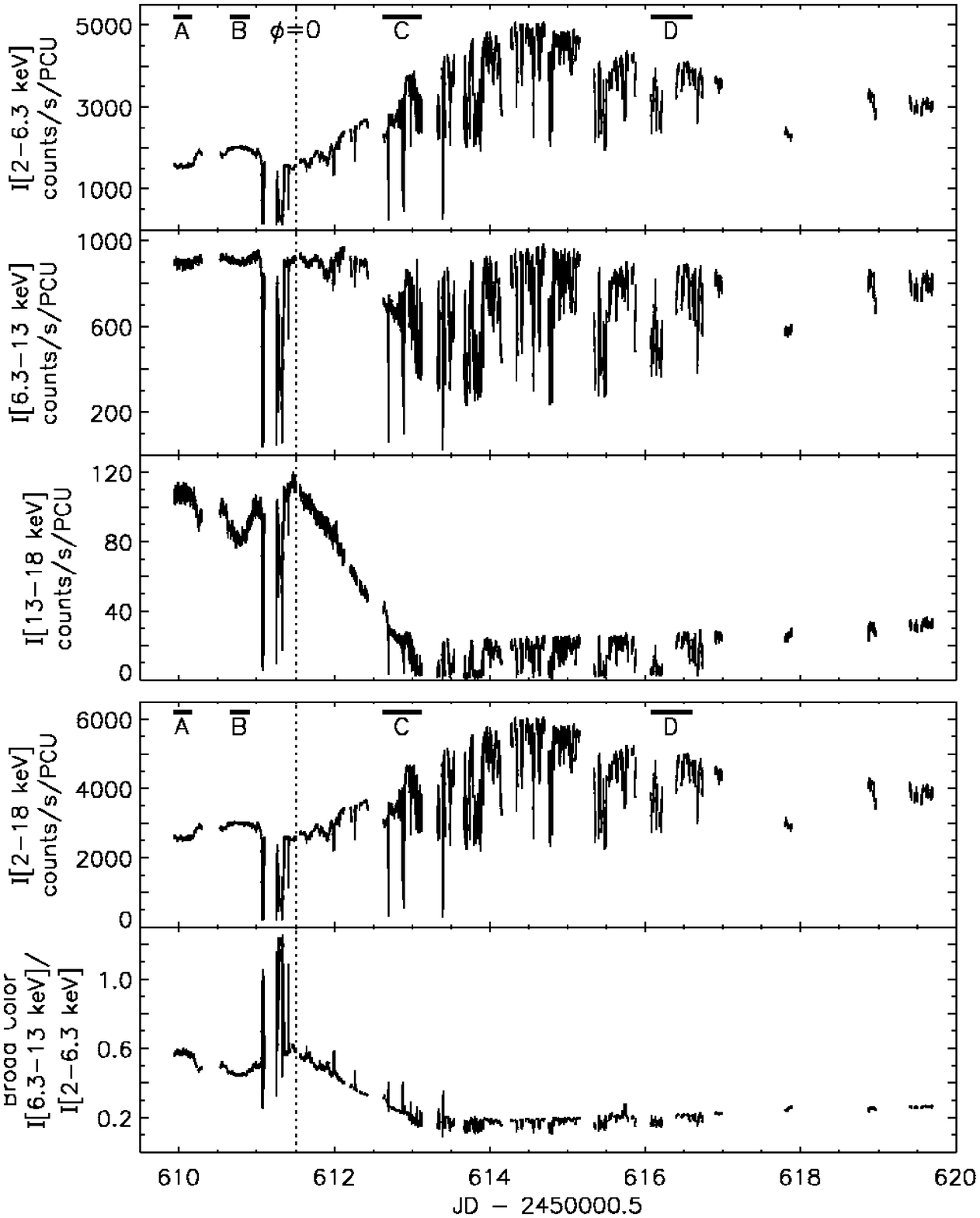}
\caption{ Light curves in three
energy bands, the full 2--18~keV band, and a hardness ratio (broad
color) for PCA observations of Cir~X-1 from 1997 June 10--20, covering
a 10-day period around phase zero ($\phi=0$; dashed
line). JD~2450609.5 = 1997 June 10 0$^h$~UT\@. The intensities at the
beginning of these observations (day~610) are typical ``quiescent''
levels.  Each point represents 16~s of background-subtracted data from
PCUs 0, 1, and 2. Segments labeled A, B, C, \& D were used for
spectral studies.
\label{fig:june97_10d}}
\end{figure}

\begin{figure}
\epsscale{1}
\plotone{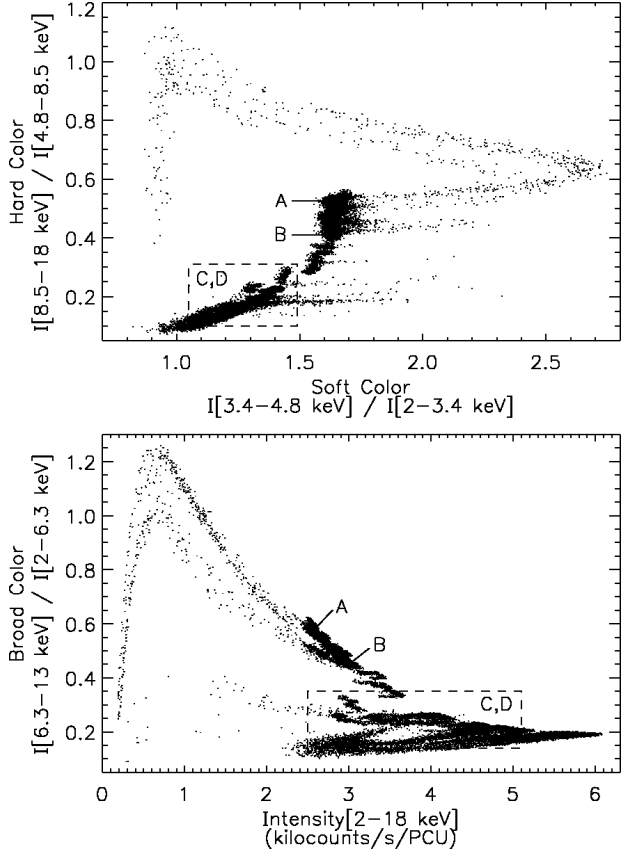}
\caption{ Color-color and
hardness-intensity diagrams from PCA observations during 1997 June
10--20 (the entire period covered by
Fig.~\protect{\ref{fig:june97_10d}}). Each point represents 16 s of
background-subtracted data from PCUs 0, 1, and 2.  Time segments A \&
B, defined in Fig.~\protect{\ref{fig:june97_10d}}, produced small
clusters whose locations are indicated in the diagrams. Time segments
C \& D produced extended tracks within the dashed boxes, which define
the plot ranges of Fig.~\protect{\ref{fig:june97_cchid_13_17}}
\label{fig:june97_cchid_all}}
\end{figure}

The color-color and hardness-intensity diagrams (CDs/HIDs) for all
data in Figure~\ref{fig:june97_10d} are shown in
Figure~\ref{fig:june97_cchid_all}. Only data from three of the five
PCA detectors, PCUs 0, 1, and 2, were included in the diagrams, since
PCUs 3 and 4 were not operating during portions of the observations.
These data cover a significant portion (10~d) of an entire 16.55~d
cycle. Data from before day 612.5 generally had soft color $\simgt$
1.5 and broad color $\simgt$ 0.325, while the data after day 612.5
generally fell below those values.

The dips seen in Figure~\ref{fig:june97_10d} appear as prominent but
less dense tracks with two sharp bends in the CD (initially toward the
right of the main arc-shaped locus) and one sharp bend in the HID
($I<2.3$~kcounts s$^{-1}$ PCU$^{-1}$).  We show in another paper
(Shirey, Levine, \& Bradt 1999\nocite{shirey99:dips}) that tracks with
these shapes are due to a variably absorbed bright spectral component
plus an unobscured faint component.  Brandt et~al.\
(1996\nocite{brandt96}) used a similar model to explain the spectral
changes during an intensity transition of Cir~X-1 observed with
\ASCA\@.  Having identified absorption dip signatures, we now focus on
non-dip spectral behavior, which is presumably more directly related
to the mechanisms of X-ray production.

Most of the data in Figure~\ref{fig:june97_cchid_all} fall along a
single arc-shaped locus in the CD and a more complicated structure in
the HID\@.  We use spectral bands with lower energies than those used
for the 1997 February--March observations in \cite{shirey98:feb97};
this enhances the branch structure in the CD\@.  When the diagrams for
the two observations are constructed in the same manner, with the same
hardness ratios and only three PCUs rather than five, the data from
these two cycles cover approximately the same extent in both
diagrams. The high-efficiency coverage of the current observations
resulted in tracks that are more complete than the tracks of the
earlier observations. Unpublished data from several other cycles
observed with the PCA also fall in similar regions of the diagrams as
the data in Figure~\ref{fig:june97_cchid_all}.

\begin{figure}
\epsscale{1}
\plotone{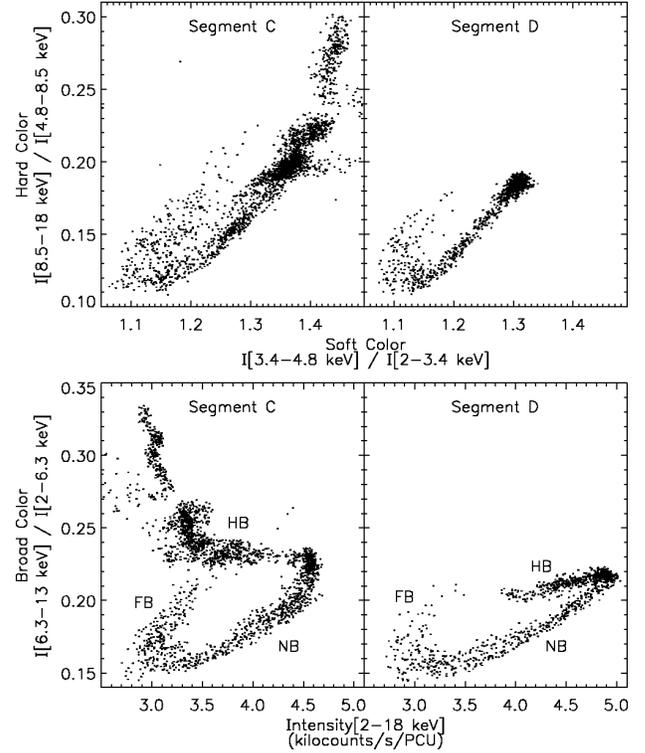}
\caption{ Color-color and hardness-intensity diagrams
from time segments C (left panels) and D (right panels) of
Fig.~\protect{\ref{fig:june97_10d}}.  In the HID, horizontal, normal,
and flaring branches (HB/NB/FB) have been identified. Each point
represents 16~s of background-subtracted data from PCUs 0, 1, and
2. The ranges of these diagrams are indicated in
Fig.~\protect{\ref{fig:june97_cchid_all}} by dashed boxes.
\label{fig:june97_cchid_13_17}}
\end{figure}

The detailed structure within the overall locus of CD/HID points is
revealed in CD/HID plots of data divided into shorter time segments
($\simlt$12~h). In particular, four time segments labelled ``A'',
``B'', ``C'', and ``D'' in Figure~\ref{fig:june97_10d} have been
selected for further study. The intensity from time segments A and B
(days 609.93--610.16 and 610.66--610.90 respectively) shows minimal
variability in each energy channel, and thus these time segments
produce small CD/HID clusters whose locations are indicated in
Figure~\ref{fig:june97_cchid_all}.
The source was more active during time segments C and D (days
612.625--613.125 and 616.075--616.600 respectively), and the CD/HID
tracks from these times show several connected branches.  Enlarged
views of the CD and HID for these time segments are shown in
Figure~\ref{fig:june97_cchid_13_17}. The full ranges of the diagrams
of Figure~\ref{fig:june97_cchid_13_17} are indicated by dashed
rectangular boxes in Figure~\ref{fig:june97_cchid_all}.
Tracks of other time segments generally each resemble some portion of
the entire pattern shown in Figure~\ref{fig:june97_cchid_13_17}, but
often with a shifted position in the diagrams. 

The data from segment~C in Figure~\ref{fig:june97_cchid_13_17} include
some absorption dips, which result in tracks moving off the right side
of the CD and the left side of the HID (and far beyond the limits of
the plot in both cases). The timing and spectral data associated with
these absorption dips will be omitted from analysis of the HID
branches.

The HID patterns reveal the shape of the full ``Z'' track as
previously inferred from the fragmented tracks in
\cite{shirey98:feb97}\@.  Time segments C and D both show a horizontal
branch, a normal branch, and a flaring branch which turns above the
normal branch. In addition, segment~C exhibits a long nearly-vertical
extension on the left end of the HB, while for segment~D, there is
only a small hint of an upward turn at the left end of the HB\@. The
HIDs show significant shifts of the HB and upper NB between the time
segments C and D, which were separated by several days.

The branches in the CDs of Figure~\ref{fig:june97_cchid_13_17} are
less well-separated than those in the HIDs. This was also the case for
the diagrams in \cite{shirey98:feb97}\@. However, the flaring branch
clearly turns above the normal branch in the lower left part of the
current CDs, and the upturned left extension of the HID horizontal
branch of segment~C is marked by an increase in the slope in the upper
right part of the associated CD.

The HID for segment~C is similar to that derived from
\RXTE\ PCA observations of the Z~source Cyg~X-2 (\cite{smale98}). The
Cyg~X-2 HID also shows a very prominent vertical extension of the
HB\@.  A similar upturned HB was reported in \Ginga\ and \EXOSAT\
observations of GX~5$-$1 (\cite{lewin92:gx5-1};
\cite{kuulkers94:gx5-1}).
The upturned flaring branch is similar to the flaring branch observed
in the CD for the Z~source GX~349+2 in recent \RXTE\ PCA observations
(\cite{zhang98}).

\begin{figure}
\epsscale{1}
\plotone{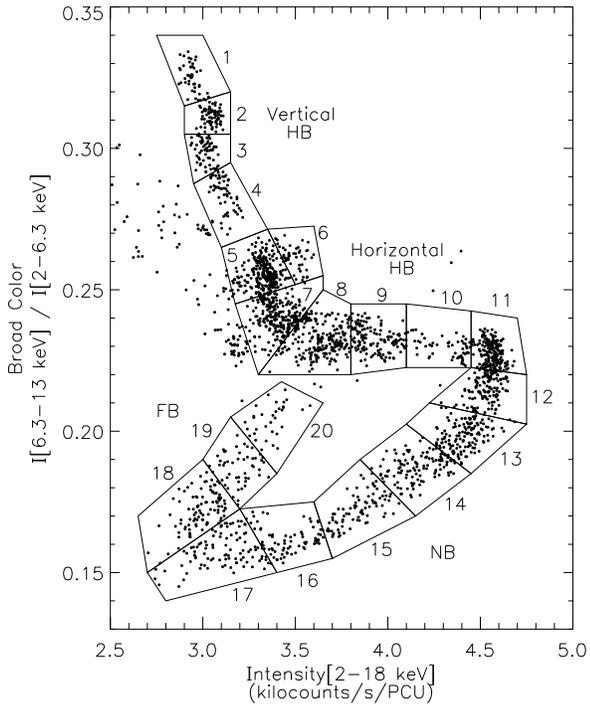}
\caption{ Hardness-intensity diagram
from time segment~C\@. The HID track has been divided into 20 regions
which were used to group data for construction of average power
density spectra and energy spectra.
\label{fig:june97_hid20reg}}
\end{figure}

The HID track for time segment~C was divided into 20 regions which
were used to group data for further timing and spectral analysis.  The
20 regions have been numbered as shown in
Figure~\ref{fig:june97_hid20reg}, with numbers increasing from the
vertical HB, through the NB, to the FB\@.  We show below that region~6
does not adhere to the otherwise monotonic variation of
spectral/temporal characteristics with region number.  This region may
be an indication of an upward-shifted horizontal portion of the HB.

\begin{figure}
\epsscale{1}
\plotone{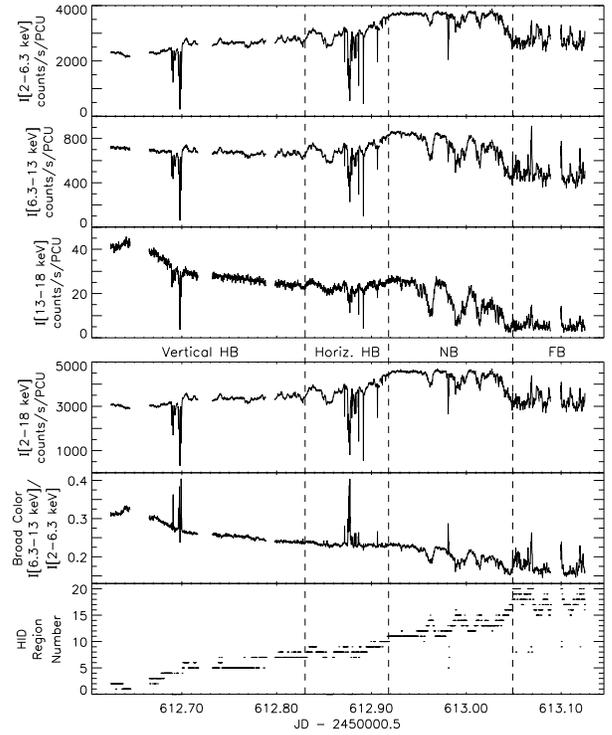}
\caption{ Top: expanded light curves
in 3 energy channels, and bottom: total 2--18~keV light curve, broad
color, and HID regions for time segment~C\@.  The predominant spectral
branch is identified for each portion of the data, based on HID region
numbers. Absorption dips (omitted from HID regions) are clearly
identified by decreased intensity coupled with pronounced increases in
broad-color.
\label{fig:june97_lc_hr_reg}}
\end{figure}

Details of the temporal variability of the intensity, hardness ratio,
and HID region numbers for time segment~C are shown in
Figure~\ref{fig:june97_lc_hr_reg}.  During this half-day segment, the
source generally moves from lower to higher region number as the
observation progresses. Thus, the time series can be divided into four
sub-segments which predominantly correspond to each portion of the HID
track: the vertical and horizontal portions of the HB, the NB, and the
FB\@. Absorption dips occur in all but the flaring branch during this
particular data set; these are easily identified by brief intensity
dips coupled with pronounced increases in broad color. No region
number was assigned to most data points from dips since the HID
regions in Figure~\ref{fig:june97_hid20reg} were selected to avoid
dips.

The light curves of the different branches
(Fig.~\ref{fig:june97_lc_hr_reg}) exhibit the following
characteristics, excluding the behavior associated with the dips.
When the source is on the vertical portion of the HB, the intensity
evolves relatively smoothly, with a slight increase in the 2--6.3~keV
band and a decrease of almost a factor of two in the 13--18~keV
band. On the horizontal portion of the HB, the source shows a
substantial increase in soft intensity and on average shows relatively
steady hard intensity. On the normal branch, the intensity is high in
the soft band while decreasing and highly variable in the hard band.
The NB/FB transition occurs at lower intensity in all bands compared
to most of the NB\@. The flaring branch itself is then produced by
high-variability ``mini-flares'' or bursts above the NB/FB apex level
(region~17).

Although the HID regions were defined such that obvious absorption
tracks were avoided, one brief dip, at day 612.98, occurred from
region 12 on the normal branch and placed a few points artificially
across regions 9, 7, and 5. These points are easily identified in
Figure~\ref{fig:june97_lc_hr_reg} (bottom plot) and are thus not
included in subsequent timing and spectral analysis.

Likewise, the highest mini-flares on the flaring branch actually
extend beyond region~20 and cross regions 8 and 9. In fact a few such
points can even be seen above region~10 in
Figure~\ref{fig:june97_hid20reg}. The FB points that fell into HB
regions can also be clearly identified as points with region numbers
of 8 or 9 in the FB portion of
Figure~\ref{fig:june97_lc_hr_reg}. These are not included in
subsequent timing and spectral analysis.

\section{Evolution of the Power Density Spectrum}

\begin{figure}
\epsscale{1}
\plotone{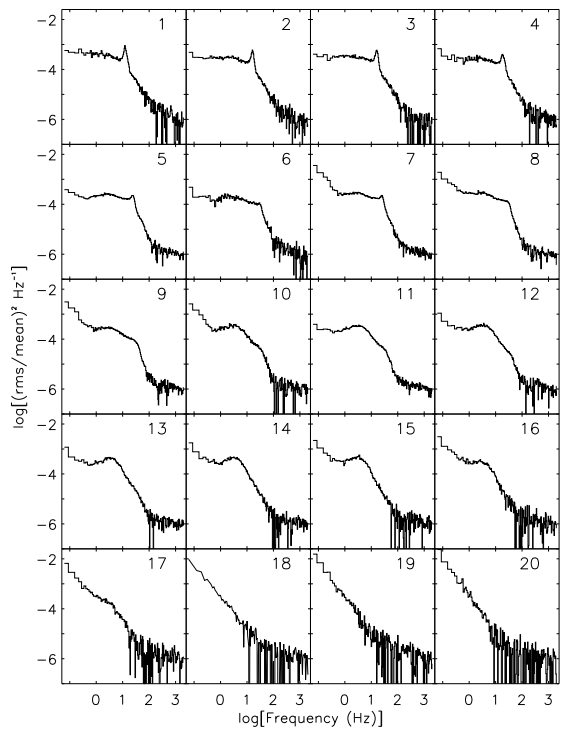}
\caption{
Averaged and rebinned Fourier power density spectra (2--32~keV) 
for each of the 20 regions along the HID track in
Fig.~\protect{\ref{fig:june97_hid20reg}}. The estimated Poisson
noise level has been subtracted from each PDS.
\label{fig:june97_20pds}}
\end{figure}

Fourier power density spectra (PDSs) were computed for each 16~s of
time segment~C\@.  Each transform used $2^{16}$ 244-$\mu$s
($2^{-12}$~s) time bins and covered the full 2--32~keV energy range.
The expected Poisson level, i.e., the level of white noise due to
counting statistics, was estimated taking into account the effects of
deadtime~(\cite{morgan97}; \cite{zhang95}; \cite{zhang96}) and
subtracted from each PDS; this method tends to slightly underestimate
the actual Poisson level. For each of the 20 HID regions defined in
Figure~\ref{fig:june97_hid20reg}, an average PDS was calculated from
the power spectra corresponding to points in that region. The average
PDSs were then logarithmically rebinned and are shown in
Figure~\ref{fig:june97_20pds}.

The general features of the power spectra are similar to those
observed in previous PCA observations (see Papers I\nocite{shirey96}
\& II\nocite{shirey98:feb97}).  In \cite{shirey98:feb97}, based on
features of the power spectra associated with fragmented spectral
tracks, we identified many of these features with those of a Z-source,
and we discussed the properties of the QPOs on the horizontal and
normal branches. In the current observations, the characteristics of
the time variability are seen to evolve smoothly during a single 12-h
observation (segment~C).  The narrow QPO is observed to evolve from
12~Hz in region~1 to 25~Hz in region~7 as the HID location moves down
the vertical extension of the HB (region 6 may be a shifted version of
8). Across the horizontal portion of the HB (regions 8--11), the
narrow QPO fades into a knee close to 30~Hz, while the broad QPO
gradually rises near 4~Hz. The broad QPO is present near 4~Hz along
the normal branch (regions 12--16). It is most prominently peaked in
the middle of the branch and weak at the bottom of the branch
(region~17). On the flaring branch (regions 18--20), no QPOs are
present and the power spectrum shows only strong very low frequency
noise.

The PDS properties of the horizontal portion of the HB are somewhat
similar to those of the upper normal branch of some Z sources, namely
no significant evolution of the HBO frequency and weak NBOs.  However,
we will continue to refer to this branch as part of the the HB since
other Z sources also show both vertical and horizontal portions of the
HB (see above).

\section{Evolution of the Energy Spectrum}
\label{sec:june97_spec_qual}

\begin{figure}
\epsscale{1}
\plotone{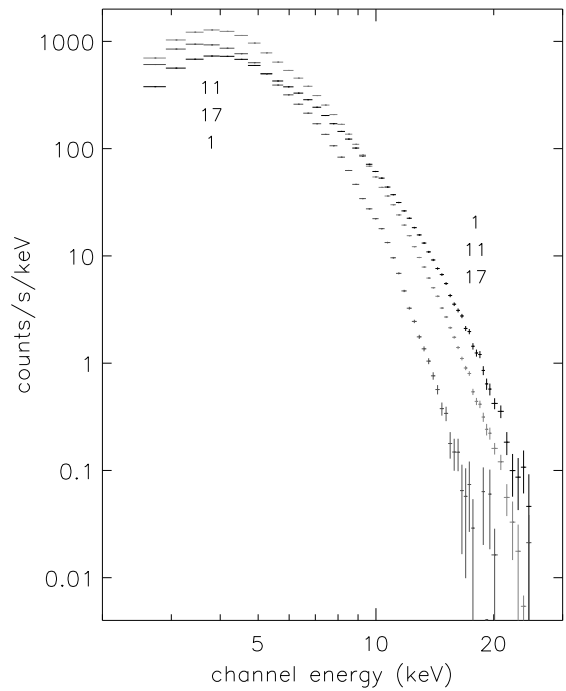}
\caption{ 
Pulse-height spectra (2.5--25~keV; PCU~0 only) from the hard, bright,
and soft extremes, regions 1, 11, and 17 respectively, of the HID
track in Fig.~\protect{\ref{fig:june97_hid20reg}}.  Labels for the
region numbers are ordered vertically to match the relative
intensities at the low and high-energy ends of the spectra.
\label{fig:june97_spec_branches}}
\end{figure}

The Standard2 data mode of the PCA instrument produces 129-channel
energy spectra every 16~s.  A parallel background file was constructed
using the ``pcabackest'' program\footnote[1]{The background model was
defined in three files provided by the PCA instrument team at
NASA/GSFC: \nl {\tt pca\_bkgd\_q6\_e03v01.mdl}, {\tt
pca\_bkgd\_xray\_e03v02.mdl}, and {\tt pca\_bkgd\_activ\_e03v03.mdl}.}
provided with the FTOOLS analysis package (version 4.0).  Average
pulse-height spectra (and background spectra) were constructed for
each of the 20 HID regions, separately for each of the five
PCUs. Version 2.2.1 response matrices were used in the analysis of
these spectra.  A 1\% systematic error estimate was added in
quadrature to the estimated statistical error (1~$\sigma$) for each
channel of the spectra to account for calibration uncertainties.
Although the instrument response matrix is imperfectly known, we can
safely assume that any spectral features that vary during the 12-hours
spanned by time segment~C are due to evolution of the source spectrum.
Representative spectra from the hard, bright, and soft extremes
(regions 1, 11, and 17, respectively) of the evolution along the HID
track are shown in Figure~\ref{fig:june97_spec_branches}.

\begin{figure}
\epsscale{1}
\plotone{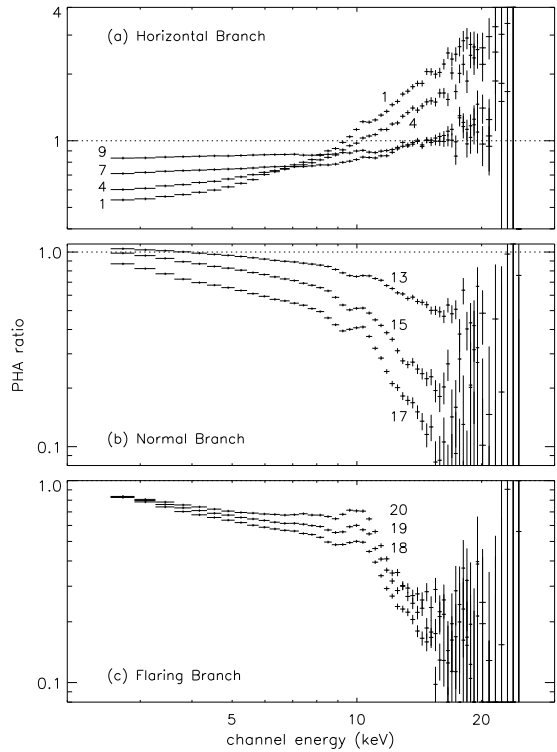}
\caption{ 
Ratio of pulse-height spectra (PCU~0 only) from selected HID regions
to the pulse-height spectrum of Region 11, showing the evolution of
the spectrum along the (a) horizontal, (b) normal, and (c) flaring
branches.
\label{fig:june97_pha_ratio}}
\end{figure}

The evolution of the spectrum may be studied by inter-comparison of
ratios of pulse-height spectra from each region to that of a reference
spectrum, from region~11 (Fig.~\ref{fig:june97_pha_ratio}).
The spectrum is hardest in region~1, at the top of the vertical
extension of the horizontal branch.  Motion down the branch
(softening, regions~1--7) corresponds to pivoting of the spectrum
about $\sim$7~keV, i.e., increasing intensity below $\sim$7~keV and
decreasing intensity above that energy.
Motion to the right across the horizontal portion of the HB
(regions~8--11) corresponds to continued increasing low-energy
intensity with modest softening, but with a nearly constant spectrum
above 12 keV\@.  Note that in the hardness ratios of
Figure~\ref{fig:june97_hid20reg} the high-energy channel (6.3--13 keV)
is dominated by photons near the lower bound of the interval.

When the source moves down the normal branch (regions 11--17), the
flux generally decreases across the entire 2.5--25~keV band but
decreases most significantly at high energy from region 11 to 15 (thus
further softening).  Moving down the NB, the spectrum gradually
develops a dip or step above $\sim$9~keV and a peak slightly above
10~keV.
Motion up the FB (regions 18--20) is produced by increasing intensity
at intermediate energies with a relatively constant spectrum at low
energy, thus hardening. The peaked feature near 10~keV becomes more
prominent moving up the FB and will be discussed below.

\section{Selection of Spectral Models}
\label{sec:june97_spec_models}

Spectral forms (e.g., blackbody emission, a power law, etc.) for use
in fitting the spectra from the HID regions were explored by first
applying them to high-quality spectra from time segments A and B (see
Fig.~\ref{fig:june97_10d}). Variability in both of these segments was
limited to less than 10\% in all energy bands between 2.5--18~keV. We
thus constructed a single (averaged) pulse-height spectrum, known
hereafter as spectrum~A or spectrum~B, for the entire 17--19~ks of
each segment.  Errors in these spectra are dominated by the 1\%
systematics at all energies up to $\sim$20~keV.

The timing properties measured throughout data sets A and B indicate
that time segment~A falls on the vertical portion of the HB (strong
narrow QPO at 8.4--11.5~Hz) and time segment~B falls near the HB/NB
apex (weak narrow QPO above 30~Hz and/or the broad 4~Hz QPO).  The
locations of time segments A and B in
Figure~\ref{fig:june97_cchid_all} indicate that the branches are
significantly shifted relative to those of segments C and D.

Remillard et~al.\ (1998\nocite{remillard98}) studied version 2.2.1 PCA
response matrices using Crab nebula data and found that the response
model was most accurate for PCUs 0, 1, and 4 (of the five PCA
detectors) and for energies between 2.5~keV and 25~keV\@. Thus, in
fitting spectra we only include data from PCUs 0, 1, and 4 and from
energy channels corresponding to 2.5--25 keV\@.  Spectra from each of
these detectors are fit separately. Fit parameters reported are the
average values for PCUs 0, 1, and, when appropriate, 4.  Errors are
conservatively estimated as the entire range encompassed by the 90\%
confidence intervals from each of the detectors. PCU~4 consistently
gives lower normalizations for fitted spectral components, so fit
parameters from that detector are not included when computing the
average normalizations and flux values and their errors.  Spectrum~B
was not constructed for PCU~4 since that detector was turned off
during part of time segment~B.

Interstellar photoelectric absorption was included in all models. The
absorption model used solar abundances (\cite{anders82}) and
cross-sections given by Morrison \& McCammon (1983\nocite{morrison83}).

Several single-component models were fit to spectra~A and
B\@. Blackbody and power-law models fit very poorly in both cases, as
did a multi-temperature ``disk blackbody'' spectrum (\cite{mitsuda84};
\cite{makishima86}; model ``diskbb'' in XSPEC), with reduced $\chi^2$
($\chi^2_r$) values of 22--545.  A thermal bremsstrahlung model
provided a better fit to spectrum~B ($\chi^2_r=4.0$), but fit
spectrum~A poorly ($\chi^2_r=34$).
A relatively good fit was achieved for both spectra with a modified
bremsstrahlung model (see Table~\ref{tab:june97_fitsAB}) which
includes the effects of Compton scattering of bremsstrahlung photons
to higher energy in an optically thick plasma cloud (\cite{compls};
model ``compLS'' in XSPEC).

A number of two-component models were also fit to these two spectra.
A model using a disk blackbody and power law did not fit well
($\chi^2_r$=3--5), mainly because a single power-law slope does
not adequately describe the spectrum at high energy.
A blackbody with $kT \simgt$ 2~keV is often included in models of the
hard X-ray emission of LMXBs thought to contain a neutron star, where
emission from or near the surface might produce high-temperature
blackbody emission with a small effective area.
Two blackbodies ($kT \sim$ 1.1~keV and 2.2~keV) fit moderately well
(see Table~\ref{tab:june97_fitsAB}), but required negligible
interstellar absorption. The low absorption is inconsistent with
previous measurements from \ASCA\ and \ROSAT\ (both sensitive below
2~keV where the absorption is most easily constrained) which were used
to estimate the interstellar column density to be
$N_H=$(1.8--2.4)$\times 10^{22}$~cm$^{-2}$
(\cite{brandt96}; \cite{predehl95}).

Two models commonly used to fit Z-source energy spectra are the
``Western model'' and the ``Eastern model'' (\cite{hasinger90};
\cite{asai94}). The Western model consists of blackbody emission from
the hot surface of the neutron star or from a boundary between the
accretion disk and surface, plus a Boltzmann-Wien component due to
unsaturated Comptonization of soft photons by hot electrons
(\cite{white86}; Schulz, Hasinger, \& Tr\"{u}mper
1989\nocite{schulz89}; Langmeier, Hasinger, \& Tr\"{u}mper
1990\nocite{langmeier90}; \cite{schulz93}).  The Eastern model also
includes blackbody radiation emitted from or near the surface, plus
emission from a multi-temperature accretion disk (\cite{mitsuda84};
\cite{hoshi91}; \cite{hirano95}).

\begin{figure}
\epsscale{1}
\plotfiddle{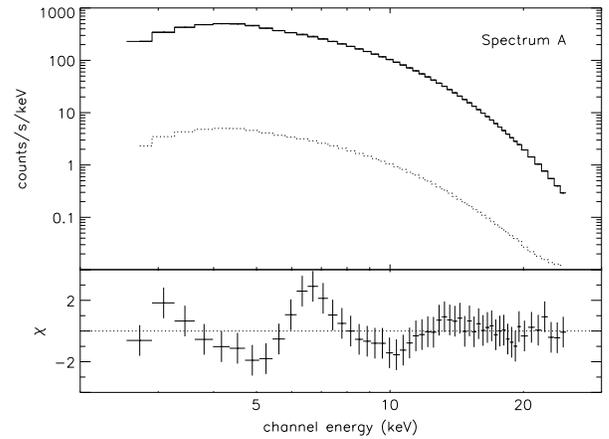}{234pt}{90}{35}{35}{140pt}{0pt}
\caption{ 
Spectrum~A and model (histogram) consisting of a disk blackbody and
blackbody (see Table~\protect{\ref{tab:june97_fitsAB}}). The
1-$\sigma$ uncertainty, shown in the top panel as a dotted curve below
the spectrum, is dominated by the 1\% systematic error for all
channels below $\sim$20~keV\@.  The residuals ($\chi$), divided by the
1$\sigma$ uncertainty, show a peak at 6--7~keV that may be due to an
iron emission line.
\label{fig:june97_specA_diskbb_bb}}
\end{figure}

The Western model, a power law with a high-energy exponential cutoff
plus a blackbody, fit well (see Table~\ref{tab:june97_fitsAB}), but
the best-fitting high-energy cutoff energy ($E_{cut}\approx1.7$~keV)
was so low relative to the PCA bandpass ($\simgt 2$~keV) that the
power law photon index was not well constrained.
The Eastern model, a disk blackbody with temperatures at the inner
edge of the disk of 1.5--1.8~keV, plus a $\sim$2~keV blackbody, fit
spectra A \& B quite well (see Table~\ref{tab:june97_fitsAB} and
Fig.~\ref{fig:june97_specA_diskbb_bb}) and gave absorption column
densities roughly consistent with the \ASCA\ and \ROSAT\ values.
Although a number of other two-component models also produce similar
quality fits, the Eastern model is used below to provide a physically
motivated parameterization of the spectra from the HID regions.

The Eastern model fit to spectrum~A
(Fig.~\ref{fig:june97_specA_diskbb_bb}) shows peaked residuals at
6--7~keV, suggesting the presence of an emission line, probably iron
K$\alpha$. Very similar residuals appear in most of the fits discussed
above for both spectra A and B\@. Addition of a Gaussian line to the
models does in fact improve the fits in almost all cases; however, the
best-fitting line often has an extremely large Gaussian width
($\sigma>1$~keV). The energy resolution of the PCA is about 1~keV FWHM
at 6~keV; thus it is difficult to place reliable constraints on
parameters such as the centroid and width of a narrow component.  We
have not included an emission line in the fits reported in
Table~\ref{tab:june97_fitsAB}. The presence of an emission line near
6.4~keV is discussed in more detail in Shirey et~al.\
(1999\nocite{shirey99:dips}) in conjunction with spectra of absorption
dips, which show the line more prominently.

\section{Fits to Spectra from 20 HID Regions}

\begin{figure}
\epsscale{1}
\plotone{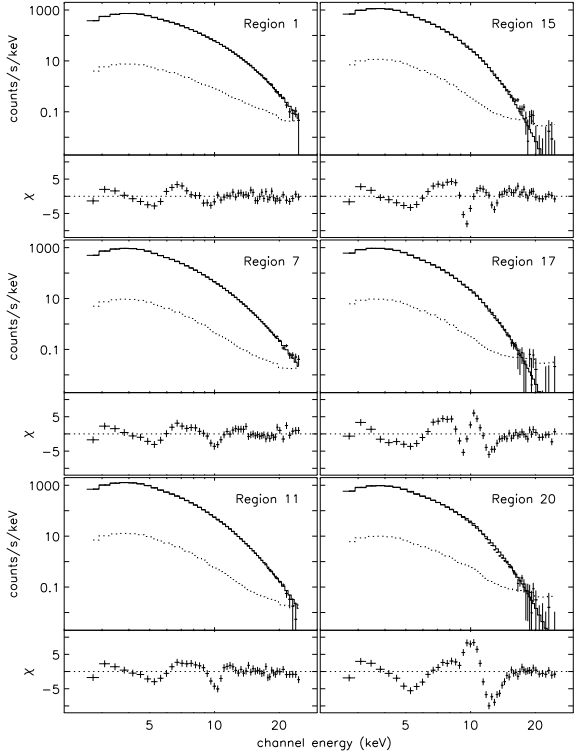}
\caption{ 
Fitted pulse-height spectra and residuals ($\chi$) for selected HID
regions (PCU~0 only).  The 1-$\sigma$ uncertainties for each spectrum,
shown as dotted curves below the spectra, are dominated by the 1\%
systematic error for all channels below $\sim$10~keV\@. Fit parameters
are listed in Table~\protect{\ref{tab:june97_fitsHID}}.  Towards
higher region number, the residuals show an unusual line- or edge-like
feature near 10~keV.
\label{fig:june97_spec_hidfits}}
\end{figure}

\begin{figure}
\epsscale{1}
\plotone{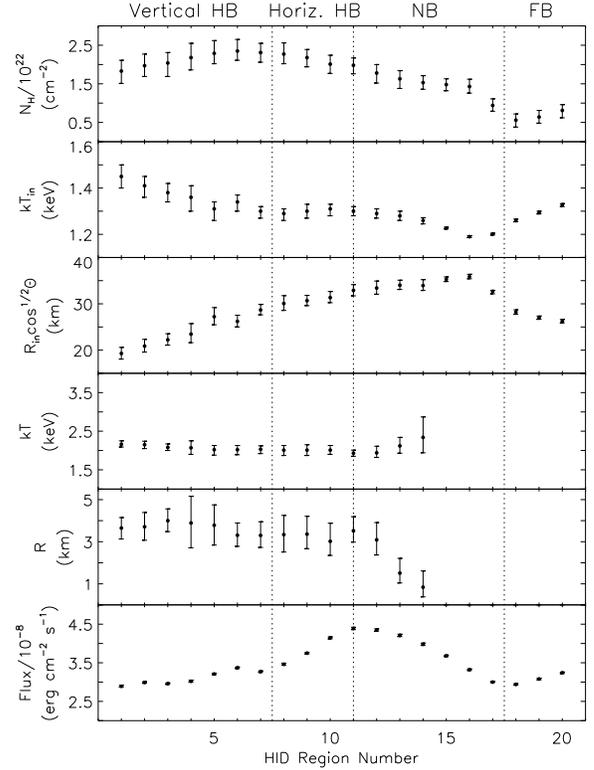}
\caption{
Spectral fit parameters for the disk blackbody plus blackbody model
versus HID region number. The plot is divided by vertical dotted lines
into four parts corresponding to branches of the HID track.
\label{fig:june97_fit_params}}
\end{figure}

A disk blackbody plus isothermal blackbody model was fit to the
average spectrum for each of the 20 HID regions.  Representative fits
and residuals are shown in Figure~\ref{fig:june97_spec_hidfits}.  The
resulting fit parameters are listed in Table~\ref{tab:june97_fitsHID}
and plotted versus HID region number in
Figure~\ref{fig:june97_fit_params}. The distance to Cir~X-1 has been
estimated to be about 6--10~kpc (\cite{stewart91}; \cite{goss77}), so
we adopt a value of 8~kpc in converting blackbody and disk blackbody
normalizations to radii.

Although both spectra A and B are fit well by the Eastern model, the
reduced chi-squared values in Table~\ref{tab:june97_fitsHID} indicate
that none of the fits for the 20 regional spectra are formally
acceptable.  The fit results must therefore be regarded as an
approximate description of the spectrum and its evolution.  We
emphasize that here, as in any case, caution is advised in drawing
physical conclusions from the best-fit model parameters.

The spectra along the horizontal branch (regions 1--11) were all fit
relatively well. The residuals for these fits (see
Fig.~\ref{fig:june97_spec_hidfits}) are similar in structure to those
for spectra A and B above. Thus they also suggest the presence of an
emission line from iron. These spectra all show column densities of
1.8--2.3$\times10^{22}$~cm$^{-2}$, consistent with the \ASCA\ and
\ROSAT\ values discussed above.  
The temperature of the $\sim$2.0~keV blackbody is relatively stable on
the HB\@ (see Fig.~\ref{fig:june97_fit_params}). The temperature of
the inner disk decreases from region 1 to region 5 (down the vertical
portion of the branch) and then stabilizes at $\sim$1.3 keV\@. The
cooling is at least in part responsible for the pivoting of the
spectrum on the vertical portion of the HB.

The inner radius of the disk blackbody component, times a factor of
order unity involving the inclination angle of the disk, increases
from 19 to 33~km, while the radius of the blackbody remains
between 3~and 4~km.  These size scales are consistent with the
hypothesis that these components arise from emission close to a
neutron star. In this model, an increasing inner radius of the
accretion disk is the most significant factor in producing the HB
track; however, one should use caution in interpreting this as an
actual physical radius.  The inclination angle of the disk in Cir~X-1
is unknown but might be high since absorption dips are observed.

From region~1 to 11, the total 2.5--25~keV flux increases
monotonically, with the exception of region~6, from
2.89$\times10^{-8}$ to 4.35$\times10^{-8}$~erg~cm$^{-2}$~s$^{-1}$,
corresponding to 1.2--1.8 times the Eddington luminosity limit for a
1.4~\msun\ neutron star at 8~kpc.

\begin{figure}
\epsscale{1}
\plotfiddle{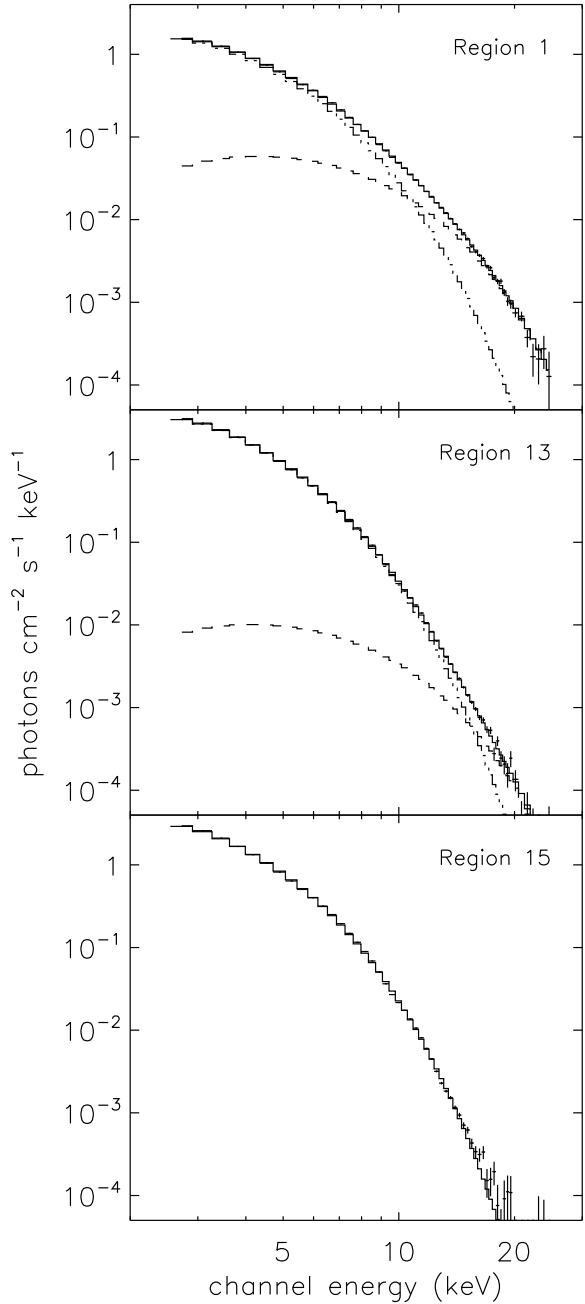}{490pt}{0}{80}{80}{-200pt}{0pt}
\caption{ 
Modeled incident spectra for HID regions 1, 13, and 15, showing the
two components (disk blackbody and blackbody) and the sum. The disk
blackbody dominates at low energy, and the blackbody generally
dominates at high energy but has faded away entirely in region~15. The
data shown are for PCU~0 only. Fit parameters are listed in
Table~\protect{\ref{tab:june97_fitsHID}}.
\label{fig:june97_photon_spec}}
\end{figure}

Along the normal branch, the quality of the fits decrease from
region~12 to region~17, as indicated by increasing $\chi^2_r$ values
(see Table~\ref{tab:june97_fitsHID}). The absorption column density
gradually decreases by a factor of two, but this may be related to the
decreasing fit quality.
The inner radius and temperature of the disk blackbody change only
slightly on the normal branch. In contrast, the $\sim$2~keV blackbody
begins to fade on the upper portion of the normal branch (regions
12--14), as indicated by a decreasing radius for the emission area.
The fading blackbody is illustrated in
Figure~\ref{fig:june97_photon_spec}, which shows the modeled incident
spectra and both model components for several spectral fits. By the
middle of the normal branch, the $\sim$2~keV blackbody has faded
entirely and fits have lower $\chi^2_r$ values without it. Thus, the
blackbody is omitted from the fits for regions 15--20. The residuals
below $\sim$6~keV continue to appear similar to those on the HB (see
Fig.~\ref{fig:june97_spec_hidfits}).  The peak at $\sim$6.5~keV
becomes broader and more complicated than on the HB\@, and the dip and
peak above 8~keV become more prominent.

On the flaring branch (regions 18--20), the fit quality decreases
further, accompanied by very low values for the absorption column
density. A number of other spectral models were fit to the HID-region
spectra, and all failed to satisfactorally fit the spectra from the
lower portion of the HID track (region number 14 and greater). A
significant contribution to the high $\chi^2$ values on the flaring
branch is due to the feature near 10~keV.  Addition of a Gaussian line
or an absorption edge at 9--11~keV does improve the fits somewhat, but
these components cannot account for all the residuals near that
energy.  A combination of a line {\em and} an edge near 10~keV can
adequately fit the residuals, but such features are difficult to
justify physically at that energy.  Even hydrogen-like iron can be
ruled out as a possible cause due to the high energy of the feature.
Many X-ray pulsars show cyclotron absorption features at high
energy. Inclusion of a cyclotron absorption component in the spectral
model results in a fit similar in quality to that of an absorption
edge. However, such features require magnetic fields of $\sim
10^{12}$~G, which would be expected to result in strong pulsations
rather than Z or atoll behavior.

\section{Discussion}

Our spectral and timing analysis of the current observations shows
clear evidence for Z~source behavior in Cir~X-1. This is significant
because Cir~X-1 was reported to exhibit atoll source behavior at lower
intensity (\cite{oosterbroek95}). Earlier \RXTE\ observations, each
lasting about two hours and separated by about two days, showed
fragments of one or two spectral branches in hardness-intensity
diagrams (\cite{shirey98:feb97})\@. In the much more extensive
observations presented in this paper, we have found longer 12-hour
segments (time intervals C and D) which clearly exhibit all three
branches of a Z~source.  We have demonstrated that these complete Z
tracks shift in the HID, confirming the behavior we inferred from the
fragmented tracks of the previous observations.

The current data also allow us to demonstrate how the timing
properties evolve along the complete HID track of Cir~X-1 and confirm
our original identification, in \cite{shirey98:feb97}\@, of horizontal
and normal branch QPOs. Fourier power spectra for different regions of
the complete HID track show continuous evolution from the narrow QPO
(increasing in frequency from 12~Hz to 30~Hz in the current
observations) on the horizontal branch, to the broad 4~Hz QPO on the
normal branch, to only very low frequency noise on the flaring branch.
Properties of the fast timing characteristics associated with spectral
branches in Cir~X-1 were discussed in \cite{shirey98:feb97}\@.  For
the remainder of this discussion we focus on the properties of the
energy spectrum.

We tried fitting energy spectra of Cir~X-1 with various simple models.
The spectra for time intervals A and B were well fit using the Western
and Eastern models (see discussion below), but no simple spectral form
was found that fit the range of spectra seen during time interval~C\@.
We have not attempted to go beyond simple parameterized spectral
models, e.g., by computing model spectra based on the "unified model"
of Lamb and collaborators (\cite{lamb89}; Psaltis, Lamb, \& Miller
1995\nocite{psaltis95}; \cite{psaltis98}), which was proposed to
explain the X-ray spectra and rapid variability of Z~sources.  Such
sophisticated models may be necessary to correctly interpret the
spectral changes in Cir~X-1 and other Z~sources.

The fits of spectra A and B with the Western model yielded a
cutoff energy of $\sim$1.7 keV for the Comptonized (Boltzmann-Wien)
component.  The cutoff energy in GX 5-1 was found to be 1--3~keV
(\cite{asai94}), similar to our results for Cir~X-1, but was found
to be higher, 4--6 keV, in Cyg~X-2 (\cite{hasinger90}).  We did not
use the Western model in parameterizing evolution associated with the
``Z'' track because the cutoff energy in Cir~X-1 is low relative to
the PCA bandpass, resulting in a poorly constrained power law index
and absorption column density.

Parameters for the Eastern model were more well-constrained (see
Table~\ref{tab:june97_fitsAB}), and thus this model was used to
parameterize the spectral variations associated with the
hardness-intensity track.  In this model, motion along the HB is
mainly associated with an increasing inner radius of the disk
(increasing disk blackbody normalization) but also by a decreasing
inner disk temperature. This would imply that, as the luminosity
increases across the HB, the inner edge of the disk is pushed further
away from the surface. It is not clear how this is related to the
increasing QPO frequency, which would typically be expected to require
a {\em decreasing} radius if the QPOs were related to Keplerian motion
at the inner edge of the disk, e.g., through the magnetospheric beat
frequency model (\cite{alpar85}; \cite{lamb85}).  

Fits of the Eastern model to energy spectra from the normal branch
indicate that the $\sim$2~keV blackbody gradually fades away, leaving
only the disk blackbody. This is similar to the result obtained when
the Eastern model was fit to the spectrum of Cyg~X-2, where the
blackbody luminosity decreases from the HB to the FB
(\cite{hasinger90}).  Furthermore, the FB of GX~5$-$1 is characterized
by intensity dips which in the Eastern model can be explained by
disappearance of the blackbody component, suggesting that accretion
flow onto the neutron-star surface is interrupted (\cite{mitsuda84}).

\begin{figure}
\epsscale{1}
\plotfiddle{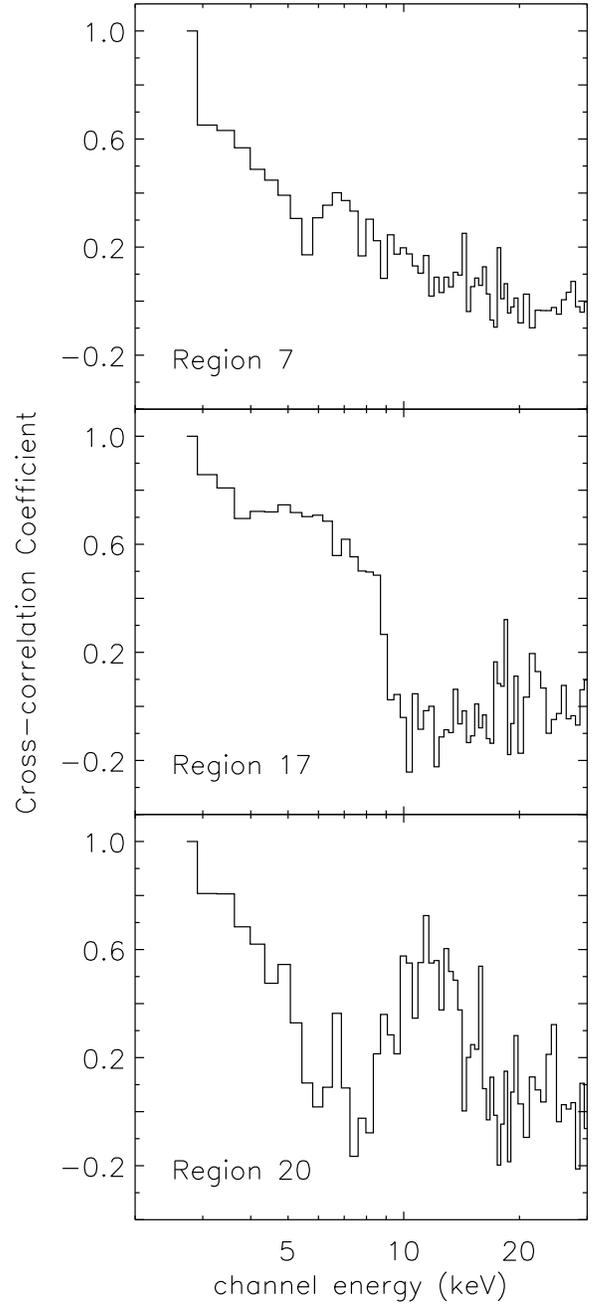}{490pt}{0}{80}{80}{-200pt}{0pt}
\caption{ 
Linear cross-correlation coefficients of time-series data (16~s bins)
in each energy channel to that of the lowest-energy channel for HID
regions 7, 17, and 20.  Towards higher region number, a step or peak
develops in the cross-correlation coefficient near 10~keV and may be
related to the feature seen at similar energy in the corresponding
pulse-height spectra and PHA ratios.
\label{fig:june97_cross_corr}}
\end{figure}

On the lower NB, a feature in the spectrum develops above 10~keV. This
feature becomes more prominent on the flaring branch.  A very similar
line-like feature at $\sim$10~keV was reported in \Ginga\ observations
of the Z~source GX~5$-$1 (\cite{asai94}). In GX~5$-$1, as in Cir~X-1,
the feature was present on the lower NB and stronger on the FB; we
thus suggest that these features of the two sources may be of similar
origin.  Asai et~al.\ showed that a peak near 11~keV occurs in the
correlation coefficients of the time-series data of different energy
bands versus the 1.7--4.0 keV band. This demonstrates a temporal
character in the narrow band at 11~keV different than that at adjacent
energies. In turn, this gives assurance that the line-like feature at
that energy is not the result of the continuum model used to fit the
spectrum but is intrinsic to the source.
We carried out similar cross-correlation analysis for each of the 20
HID regions in our study. The cross-correlation results, relative to
the 2.5--2.9~keV band, from three representative regions are shown in
Figure~\ref{fig:june97_cross_corr}.  We find a clear peak in the
cross-correlation coefficient at about 11~keV in regions 19 and 20 of
the FB\@, further confirming the similarity of the spectral features
in Cir X-1 and GX~5$-$1.  Regions 16 and 17 of the lower NB show an
abrupt drop in the cross-correlation coefficient above 8--10~keV.  The
cross-correlation coefficients on the HB and upper NB, where the line-
or edge-like 10~keV spectral feature is absent or weak, generally show
much less remarkable behavior.

Asai et al.\ found that that the spectral feature near 10~keV in
GX~5$-$1 was better fit with a Gaussian line rather than, e.g., an
absorption edge.  However, as we mentioned above, such a feature
cannot be produced by even hydrogen-like iron. Asai et~al.\ discuss
several of the mechanisms that could possibly produce a line at
$\sim$10~keV. For example, emission from a heavy element such as Ni
could produce the line, but is unlikely since iron should be far more
abundant than the heavier elements. Alternately, a line could be
blue-shifted due to motion in a relativistic jet or rotation in the
accretion disk, but extreme conditions would be required to boost the
energy up to $\sim$10~keV and one might expect a red-shifted line to
also be observed in the X-ray spectrum.  As mentioned above, we find
that in Cir~X-1 an emission line component alone is insufficient to
produce the observed structure of the feature. Both line-like and
edge-like components may be required to explain this unusual spectral
feature at high energy.

Cir X-1 exhibits a number of unusual features in its Z-source behavior
in addition to the 10~keV spectral feature.  Its HBOs are observed at
frequencies as low as 1.3~Hz (\cite{shirey96}), an order of magnitude
lower than those of other Z~sources.  The highest frequency reached by
HBOs in Cir X-1, at the HB/NB apex, is 30--32~Hz
(\cite{shirey98:feb97}), about a factor of two below the extreme HBO
frequency in other Z sources.  Power spectra of Cir~X-1 show a broad
high-frequency peak, centered at 20--200~Hz, which shifts in frequency
maintaining an almost constant ratio with the HBO frequency
(\cite{shirey96}; \cite{tennant87}).  Cir X-1 shows a long vertical
extension of the HB\@.  The entire HID track shows very large
color/intensity shifts, possibly associated with the presumed
eccentric 16.55-d binary orbit (\cite{shirey98:feb97}).  Atoll
behavior has been reported at lower intensity (\cite{oosterbroek95}).
It is likely that some or many of these unusual features are related
by some physical property of the system, indicating that Cir~X-1 may
provide important constraints on models of low-mass X-ray binaries.

\acknowledgements 

We would like to acknowledge the \RXTE\ teams at MIT and GSFC for
their support.  In particular we thank E.~Morgan, R.~Remillard,
W.~Cui, and D.~Chakrabarty for useful discussions related to this
work.  We thank D.~Psaltis for useful discussions regarding Z-source
spectral models. We also appreciate the detailed comments and
suggestions provided by the referee.  Support for this work was
provided through NASA Contract NAS5-30612.


\newpage

\newpage

\newpage

\newpage

\newpage

\newpage

\newpage

\newpage

\newpage

\newpage

\newpage

\newpage

\newpage

\newpage

\begin{deluxetable}{ccccccccc}
\footnotesize
\renewcommand{\arraystretch}{1.5}
\tablewidth{0pt}
\tablecaption{ \label{tab:june97_fitsAB}
Fit parameters for spectra~A \& B for four models.
}
\tablehead{
\colhead{} &
\colhead{$N_H/10^{22}$}  &
\colhead{} &
\colhead{} &
\colhead{} &
\colhead{} & 
\colhead{} &
\colhead{Flux/10$^{-8}$~\tablenotemark{b}} &
\colhead{} 
\nl
\phm{} & (cm$^{-2}$) & \multicolumn{5}{c}{Model Components\tablenotemark{a}} 
& erg\,cm$^{-2}$\,s$^{-1}$ & $\chi^2_r$~\tablenotemark{c} 
}
\startdata
\phm{} & \phm{} & \phm{} & \multicolumn{3}{c}{Self-Comptonized Bremsstrahlung} & \phm{} & \phm{} & \phm{} \nl
\cline{4-6}
\phm{} & \phm{} & \phm{} &  $kT$   & Optical & \phm{} & \phm{} & \phm{} & \phm{} \nl
\phm{} & \phm{} & \phm{} & (keV)  & depth   & norm & \phm{} & \phm{} & \phm{} \nl
\cline{3-7}
A & $2.80^{+0.22}_{-0.23}$ & \phm{} 
  & $2.78^{+0.04}_{-0.05}$ & $11.05^{+0.47}_{-0.36}$ & $4.54^{+0.24}_{-0.23}$ & 
  \phm{} & $2.73^{+0.02}_{-0.02}$ & 1.16--1.55 \nl
B & $3.94^{+0.18}_{-0.21}$ & \phm{} 
  & $2.89^{+0.12}_{-0.11}$ & $7.20^{+0.71}_{-0.69}$ & $9.67^{+0.37}_{-0.43}$ & 
  \phm{} & $3.04^{+0.02}_{-0.02}$ & 1.51--2.02 \nl
\hline
 & & \multicolumn{2}{c}{Blackbody} & & \multicolumn{2}{c}{Blackbody} & \nl
\cline{3-4} \cline{6-7} 
\phm{} & \phm{} & $kT$   & $R$~\tablenotemark{d}  & \phm{} & $kT$   & $R$~\tablenotemark{d} & \phm{} & \phm{} \nl 
\phm{} & \phm{} & (keV) & (km) & \phm{} & (keV) & (km) & \phm{} & \phm{} \nl
\cline{3-7}
A & $0.00^{+0.02}_{-0.00}$ 
  & $1.16^{+0.02}_{-0.02}$ & $24.13^{+0.38}_{-0.37}$ & \phm{} 
  & $2.33^{+0.02}_{-0.03}$ & $5.38^{+0.16}_{-0.15}$ 
  & $2.72^{+0.02}_{-0.02}$ & 2.51--2.82 \nl
B & $0.01^{+0.21}_{-0.01}$ 
  & $1.12^{+0.01}_{-0.02}$ & $30.46^{+0.92}_{-0.40}$ & \phm{} 
  & $2.20^{+0.03}_{-0.03}$ & $5.59^{+0.27}_{-0.25}$ 
  & $3.04^{+0.02}_{-0.02}$ & 1.42--1.78 \nl
\hline
\phm{} & \phm{} & \multicolumn{3}{c}{Cutoff Power Law~\tablenotemark{e}} & \multicolumn{2}{c}{Blackbody} & \phm{} & \phm{} \nl
\cline{3-5} \cline{6-7} 
\phm{} & \phm{} & Photon & $E_{cut}$ & \phm{} &  $kT$  & $R$~\tablenotemark{d}  & \phm{} & \phm{} \nl
\phm{} & \phm{} & index  &   (keV)   & norm & (keV) & (km) & \phm{} & \phm{} \nl
\cline{3-7}
A &  $0.27^{+0.94}_{-0.27}$ 
  & $-0.65^{+0.72}_{-0.30}$ & $1.73^{+0.58}_{-0.24}$ & $2.24^{+1.53}_{-0.47}$ 
  & $2.44^{+0.10}_{-0.05}$ & $4.40^{+0.34}_{-0.87}$ 
  & $2.73^{+0.02}_{-0.02}$ & 0.79--1.04 \nl
B & $1.95^{+0.74}_{-0.86}$ 
  & $-0.34^{+0.46}_{-0.60}$ & $1.67^{+0.31}_{-0.27}$ & $5.97^{+2.70}_{-2.29}$ 
  & $2.29^{+0.09}_{-0.07}$ & $4.61^{+0.67}_{-0.73}$ 
  & $3.04^{+0.02}_{-0.02}$ & 0.79--1.35 \nl
\hline
 & & \multicolumn{2}{c}{Disk Blackbody} & & \multicolumn{2}{c}{Blackbody} & \nl
\cline{3-4} \cline{6-7} 
\phm{}  & \phm{} & $kT_{in}$ & $R_{in}\cos^{1/2}\theta$~\tablenotemark{f} & \phm{} &  $kT$  & $R$~\tablenotemark{d}  & \phm{} & \phm{} \nl
\phm{}  & \phm{} &  (keV)   &  (km)                    & \phm{} & (keV) & (km) & \phm{} & \phm{} \nl
\cline{3-7}
A & $1.44^{+0.26}_{-0.27}$ 
  & $1.81^{+0.08}_{-0.06}$ & $9.39^{+0.64}_{-0.64}$ & \phm{} 
  & $2.47^{+0.04}_{-0.04}$ & $4.10^{+0.25}_{-0.25}$ 
  & $2.73^{+0.02}_{-0.02}$ & 0.79--1.06 \nl
B & $2.49^{+0.23}_{-0.23}$ 
  & $1.54^{+0.04}_{-0.04}$ & $15.63^{+0.93}_{-0.84}$ & \phm{} 
  & $2.28^{+0.05}_{-0.05}$ & $4.66^{+0.37}_{-0.37}$  
  & $3.04^{+0.02}_{-0.02}$ & 0.75--1.31 \nl
\enddata
\tablenotetext{a}{
Errors quoted are 90\% confidence limits for a single parameter
($\Delta\chi^2=2.7$).}
\tablenotetext{b}{Total 2.5--25~keV flux.}
\tablenotetext{c}{
$\chi^2/dof$, where $dof =$ the
number of spectral bins (52--54 per spectrum) minus the number of fit
parameters (4--6). The range of values represents the fits for the several 
PCU detectors.}
\tablenotetext{d}{Blackbody radius for a distance of 8~kpc.}
\tablenotetext{e}{Power law with exponential cut-off above $E_{cut}$ (a Boltzmann-Wien spectrum).}
\tablenotetext{f}{
Inner radius of the accretion disk (times
$\cos^{1/2}\theta$, where $\theta$ is the angle between the normal to
the disk and the line of sight) for a distance of 8~kpc.}
\renewcommand{\arraystretch}{1}
\end{deluxetable}

\begin{deluxetable}{cccccccc}
\footnotesize
\renewcommand{\arraystretch}{1.5}
\tablecaption{ \label{tab:june97_fitsHID}
Fit parameters for HID regions 1--20 using a model consisting of a 
disk~blackbody plus a blackbody.
}
\tablehead{
\colhead{HID~\tablenotemark{a}} & \colhead{$N_H/10^{22}$} & \colhead{$kT_{in}$} & \colhead{$R_{in}\cos^{1/2}\theta$~\tablenotemark{b}} & \colhead{$kT$} & \colhead{$R$~\tablenotemark{c}} & \colhead{Flux/10$^{-8}$~\tablenotemark{d}} & \colhead{} \nl 
\colhead{region} & \colhead{(cm$^{-2}$)} & \colhead{(keV)} & \colhead{(km)}
                 & \colhead{(keV)} & \colhead{(km)} & \colhead{erg\,cm$^{-2}$\,s$^{-1}$} & \colhead{$\chi^2_r$~\tablenotemark{e}} 
}
\startdata
     1 & $1.83^{+0.28}_{-0.32}$ & $1.45^{+0.05}_{-0.05}$ & $19.29^{+1.31}_{-1.22}$ & $2.16^{+0.09}_{-0.07}$ & $3.65^{+0.50}_{-0.52}$ & $2.89^{+0.03}_{-0.02}$ & 1.41--2.14 \nl
     2 & $1.97^{+0.30}_{-0.28}$ & $1.41^{+0.04}_{-0.05}$ & $20.87^{+1.46}_{-1.29}$ & $2.15^{+0.09}_{-0.10}$ & $3.71^{+0.68}_{-0.64}$ & $2.99^{+0.02}_{-0.02}$ & 1.40--1.70 \nl
     3 & $2.04^{+0.27}_{-0.35}$ & $1.38^{+0.04}_{-0.04}$ & $22.24^{+1.30}_{-1.16}$ & $2.08^{+0.09}_{-0.08}$ & $4.00^{+0.56}_{-0.52}$ & $2.96^{+0.02}_{-0.02}$ & 1.32--1.81 \nl
     4 & $2.18^{+0.37}_{-0.32}$ & $1.36^{+0.05}_{-0.06}$ & $23.46^{+2.27}_{-1.85}$ & $2.07^{+0.18}_{-0.17}$ & $3.89^{+1.27}_{-1.18}$ & $3.02^{+0.03}_{-0.02}$ & 2.21--3.47 \nl
     5 & $2.29^{+0.33}_{-0.27}$ & $1.31^{+0.03}_{-0.05}$ & $27.22^{+1.97}_{-1.74}$ & $2.02^{+0.11}_{-0.14}$ & $3.78^{+0.97}_{-0.94}$ & $3.21^{+0.02}_{-0.02}$ & 2.51--3.24 \nl
     6 & $2.35^{+0.30}_{-0.24}$ & $1.34^{+0.03}_{-0.04}$ & $26.21^{+1.33}_{-1.19}$ & $2.02^{+0.11}_{-0.13}$ & $3.31^{+0.58}_{-0.53}$ & $3.37^{+0.02}_{-0.02}$ & 1.41--1.55 \nl
     7 & $2.31^{+0.24}_{-0.25}$ & $1.30^{+0.02}_{-0.03}$ & $28.67^{+1.20}_{-1.05}$ & $2.03^{+0.09}_{-0.11}$ & $3.30^{+0.65}_{-0.57}$ & $3.27^{+0.02}_{-0.02}$ & 2.22--2.76 \nl
     8 & $2.27^{+0.29}_{-0.25}$ & $1.29^{+0.02}_{-0.03}$ & $30.06^{+1.70}_{-1.48}$ & $2.01^{+0.12}_{-0.14}$ & $3.34^{+0.91}_{-0.83}$ & $3.46^{+0.03}_{-0.02}$ & 2.41--2.70 \nl
     9 & $2.18^{+0.21}_{-0.24}$ & $1.30^{+0.03}_{-0.03}$ & $30.68^{+1.15}_{-1.07}$ & $2.01^{+0.14}_{-0.14}$ & $3.36^{+0.85}_{-0.69}$ & $3.75^{+0.02}_{-0.02}$ & 2.29--2.44 \nl
    10 & $2.01^{+0.23}_{-0.24}$ & $1.31^{+0.02}_{-0.03}$ & $31.35^{+1.31}_{-1.10}$ & $2.01^{+0.12}_{-0.11}$ & $3.02^{+0.86}_{-0.67}$ & $4.15^{+0.02}_{-0.03}$ & 1.82--2.56 \nl
    11 & $1.98^{+0.19}_{-0.22}$ & $1.30^{+0.02}_{-0.02}$ & $32.90^{+1.24}_{-1.17}$ & $1.93^{+0.08}_{-0.08}$ & $3.52^{+0.67}_{-0.54}$ & $4.39^{+0.03}_{-0.03}$ & 2.55--3.05 \nl
    12 & $1.78^{+0.22}_{-0.26}$ & $1.29^{+0.02}_{-0.02}$ & $33.41^{+1.51}_{-1.34}$ & $1.94^{+0.17}_{-0.12}$ & $3.09^{+0.82}_{-0.72}$ & $4.35^{+0.03}_{-0.03}$ & 2.87--3.06 \nl
    13 & $1.63^{+0.21}_{-0.25}$ & $1.28^{+0.02}_{-0.02}$ & $34.02^{+1.09}_{-0.92}$ & $2.12^{+0.22}_{-0.19}$ & $1.51^{+0.70}_{-0.47}$ & $4.21^{+0.03}_{-0.03}$ & 3.74--4.66 \nl
    14 & $1.53^{+0.18}_{-0.17}$ & $1.26^{+0.01}_{-0.02}$ & $33.94^{+1.28}_{-1.06}$ & $2.34^{+0.53}_{-0.40}$ & $0.84^{+0.77}_{-0.46}$ & $3.98^{+0.03}_{-0.03}$ & 4.19--5.11 \nl
    15 & $1.48^{+0.15}_{-0.16}$ & $1.22^{+0.01}_{-0.01}$ & $35.30^{+0.51}_{-0.49}$ & \nodata & \nodata & $3.68^{+0.02}_{-0.02}$ & 5.15--6.34 \nl
    16 & $1.43^{+0.19}_{-0.17}$ & $1.19^{+0.01}_{-0.01}$ & $35.88^{+0.50}_{-0.48}$ & \nodata & \nodata & $3.32^{+0.02}_{-0.02}$ & 5.91--7.52 \nl
    17 & $0.94^{+0.17}_{-0.15}$ & $1.20^{+0.01}_{-0.01}$ & $32.69^{+0.23}_{-0.53}$ & \nodata & \nodata & $3.00^{+0.02}_{-0.01}$ & 5.87--7.16 \nl
    18 & $0.56^{+0.16}_{-0.18}$ & $1.26^{+0.01}_{-0.01}$ & $28.22^{+0.49}_{-0.48}$ & \nodata & \nodata & $2.94^{+0.02}_{-0.02}$ & 9.59--11.39 \nl
    19 & $0.64^{+0.17}_{-0.16}$ & $1.29^{+0.01}_{-0.01}$ & $27.02^{+0.35}_{-0.36}$ & \nodata & \nodata & $3.08^{+0.02}_{-0.02}$ & 9.36--10.45 \nl
    20 & $0.81^{+0.15}_{-0.19}$ & $1.33^{+0.01}_{-0.01}$ & $26.25^{+0.38}_{-0.37}$ & \nodata & \nodata & $3.24^{+0.02}_{-0.02}$ & 14.67--17.01 \nl
\enddata
\scriptsize
\tablenotetext{a}{
Errors quoted are 90\% confidence limits for a single parameter
($\Delta\chi^2=2.7$).}
\tablenotetext{b}{
Inner radius of the accretion disk (times $\cos^{1/2}\theta$, where
$\theta$ is the angle between the normal to the disk and the line of
sight) for a distance of 8~kpc.}
\tablenotetext{c}{Blackbody radius for a distance of 8~kpc.}
\tablenotetext{d}{Total 2.5--25~keV flux.}
\tablenotetext{e}{
$\chi^2/dof$, where $dof =$ the number of spectral bins (52--54 per spectrum) 
minus the number of fit parameters (3--5). The range of values represents the fits for the several 
PCU detectors.}
\renewcommand{\arraystretch}{1}
\end{deluxetable}


\end{document}